\documentclass[
    ,final            
  ]
  {aipproc}

\layoutstyle{6x9}

\usepackage{amssymb}
\newcommand{\ud}{\mathrm{d}}

\begin{document}

\title{The study of quark-gluon matter in high-energy nucleus-nucleus collisions}



\author{A. Andronic}{
  address={Reasearch Division and EMMI, GSI Helmholtzzentrum f\"ur Schwerionenforschung,
D-64291 Darmstadt, Germany}
}

\begin{abstract}
A short overview is given on the study of hot matter produced in relativistic 
nucleus-nucleus collisions, with emphasis on recent measurements at the LHC.
\end{abstract}

\maketitle

The goal of high-energy nucleus-nucleus collisions is to produce and 
characterize a state of nuclear (QCD) matter at (energy) densities well above 
the nuclear ground state ($\varepsilon_0\simeq$0.15 GeV/fm$^3$).
At high densities and/or at high temperatures one expects 
\cite{Collins:1974ky,Cabibbo:1975ig}
that quarks are no longer confined in protons and neutrons but move
freely over distances larger than the size of the nucleon 
($\simeq$1 fm=10$^{-15}$ m).
Such a deconfined state of matter, earlier named the Quark-Gluon 
Plasma (QGP) \cite{Shuryak:1978ij}, was the state of the Universe within 
the first ($\simeq$10) microseconds of its creation in the Big Bang 
\cite{Boyanovsky:2006bf} and may exist as well in the core of neutron stars.
The characterization of quark-gluon matter in terms of its equation of state 
(EoS, relating pressure to energy) and of its transport properties (like 
viscosity) and delineating its phase diagram \cite{BraunMunzinger:2008tz} 
is a major ongoing research effort. 
See \cite{Satz:2012zz} for a recent extended treatment of the topic.

At low energies (beam energies per nucleon of up to 10 GeV/$A$ on fixed 
target, corresponding to center of mass energies per nucleon pair, 
$\sqrt{s_{\mathrm{NN}}}\lesssim$5 GeV) it is expected that compressed nucleonic 
matter is produced. 
The EoS of nuclear matter \cite{Danielewicz:2002pu} at densities a few times
normal nuclear density ($\rho_0=0.17$ fm$^{-3}=2.7\cdot 10^{14}$ g/cm$^3$), 
expressed as the nuclear compressibility, has relevance for the maximum 
mass of neutron stars (see ref. \cite{Klahn:2011fb} for a recent overview).

Employing Quantum Chromo-Dynamics calculations on lattice, a deconfinement 
phase transition for an energy density of about 1 GeV/fm$^3$ was 
predicted (see \cite{Karsch:2001cy} for an early review).
It was shown \cite{Aoki:2006we} that the phase transition at zero 
baryochemical potential, $\mu_b$, is of crossover type, namely with a
continuous, smooth, increase of thermodynamic quantities.
The value of the \mbox{(pseudo-)}critical temperature, $T_c$, at vanishing 
baryochemical potential ($\mu_b$) is currently estimated to be in the
range 155-160 MeV \cite{Aoki:2009sc,Bazavov:2011nk}.
The existence of a critical point (denoting the end of the first order 
phase transition line, a point where the phase transition is of a second 
order) is a fundamental question, addressed both experimentally 
\cite{Aggarwal:2010cw} and theoretically \cite{Philipsen:2011zx}.

A nucleus-nucleus collision is a highly dynamical event.
One can identify, schematically, the following stages of the system 
(``fireball''):
i) initial collisions, occuring during the passing time of the nuclei
($t_{pass}=2R/\gamma_{\mathrm{cm}} c$);
ii) thermalization: equilibrium is established;
iii) expansion and cooling (in a deconfined state);
iv) chemical freeze-out (possibly at hadronization): inelastic collisions 
cease, hadron yields (and distribution over species) are frozen;
v) kinetic freeze-out: elastic collisions cease, spectra and correlations 
are frozen.

The challenge is to characterize the hot (deconfined) stage iii), while
most of the measurements are performed via hadrons (or their decay products) 
carrying information from the system at stages iv) and v).
Even though the early stage of hot deconfined matter remains inaccessible in a 
direct way because of quark confinement, there are experimental observables 
which carry information from this stage.
Extracting the properties of the deconfined stage is possible only via models.
At lower energies hadronic transport models \cite{Danielewicz:2002pu} are 
employed, while at higher energies hydrodynamics \cite{Huovinen:2006jp} 
is widely used.
Based on model comparison to data, one can extract the following ranges of 
the fireball characteristics:
{\it Temperature:} $T=100-1000$ MeV, or up to a million times the 
temperature at the center of the Sun (\mbox{1 MeV$\simeq$10$^{10}$ K});
{\it Pressure:} $P=100-300$ MeV/fm$^3$ (1 MeV/fm$^3 \simeq 10^{33}$ Pa); 
{\it Density:} $\rho=1-10\cdot\rho_0$;
{\it Volume:} several thousands fm$^3$; 
{\it Duration:} 10-20 fm/$c$ (or about $3-6\cdot 10^{-23}$ s).

The experimental  ``control parameters'' are:
a) the collision energy (per nucleon pair, $\sqrt{s_{\mathrm{NN}}}$);
b) the centrality (impact parameter, $b$) of the collision (or, alternatively,
the size of the colliding nuclei), which is deduced from specific measurements 
and involves calculations within the Glauber model \cite{Miller:2007ri}.
A usual way of expressing centrality is via the number of participating 
nucleons, $N_{part}$, namely the nucleons involved in the creation of the 
fireball in the overlap region of the two colliding nuclei.

After the initial measurements at the Bevalac (Berkeley) in the '80s, 
the program of heavy-ion collisions continued at higher energies at 
Brookhaven at the Alternating Gradient Synchrotron (AGS) and at CERN 
at the Super-Proton Synchrotron (SPS), while in the low energy range 
measurements were performed at GSI Darmstadt at the 
Schwerionensynchrotron (SIS).
Started in year 2000, the experimental program at the Relativistic Heavy 
Ion Collider (RHIC) at Brookhaven spans $\sqrt{s_{\mathrm{NN}}}$ from 
$\sim$8 to 200 GeV (see, for earlier results, experimental summaries in 
\cite{Arsene:2004fa,Adcox:2004mh,Back:2004je,Adams:2005dq} and an overview
in \cite{Muller:2006ee}).
The study of QCD matter has entered a new era in year 2010 with the 
advent of Pb--Pb collisions at the Large Hadron Collider (LHC), 
delivering the largest ever collision energy, $\sqrt{s_{\mathrm{NN}}}$=2.76 TeV,
more than a factor of 10 larger than previously available. 
A recent overview of the first LHC data is available in 
ref. \cite{Muller:2012zq}.

\begin{figure}[htb]
\begin{tabular}{lr} \begin{minipage}{.52\textwidth}
\centerline{\includegraphics[width=.94\textwidth]{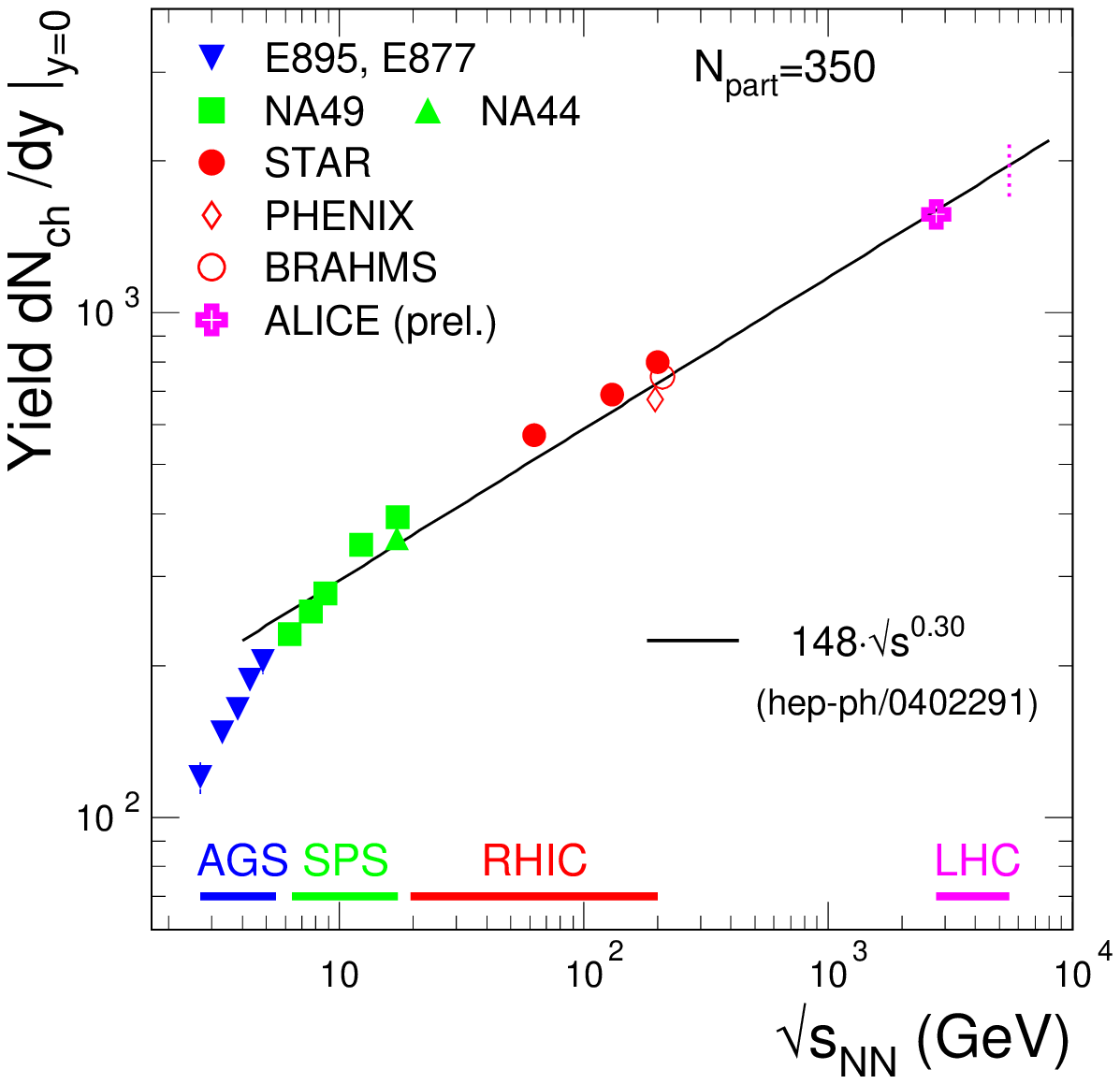}}
\end{minipage} & \begin{minipage}{.47\textwidth}
\vspace{-2.9cm}
\centerline{\includegraphics[width=1.15\textwidth]{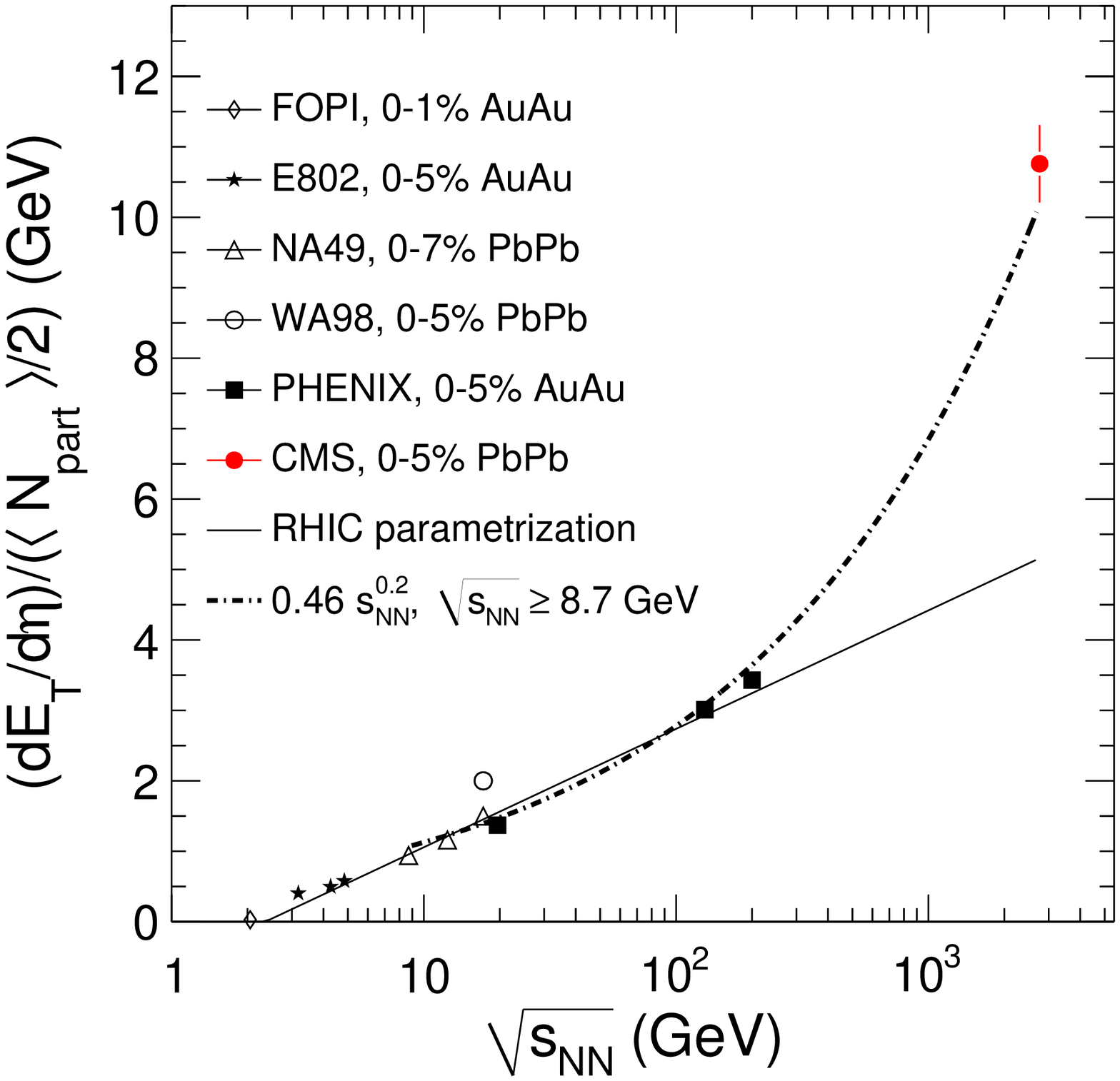}}
\end{minipage}\end{tabular}
\caption{Left panel: collision energy dependence of charged particle density 
$\ud N_{\mathrm{ch}}/\ud y$ at midrapidity, measured by various experiments in 
central collisions corresponding to $N_{part}$=350.
Right panel: energy dependence of the transverse energy 
(plot from \cite{Chatrchyan:2012mb}).
} 
\label{fig:dnchdy} 
\end{figure}

The measurement of the charged hadrons pseudo-rapidity 
($\eta=-\ln[\tan(\theta/2)]$, with $\theta$ the polar emmission angle)
density, $\ud N_{\mathrm{ch}}/\ud\eta$, at the LHC \cite{Aamodt:2010pb} was 
eagerly awaited and showed that the increase compared to the measurement 
at RHIC is by a factor of about 2.2 for central collisions. Interpreted
as the outcome of an increase of the initial entropy density, this increase
can be translated into a factor of 1.3 increase of the initial temperature
\cite{Physics.3.105}. 
The ALICE measurement confirmed the phenomenological 
$(\sqrt{s_{\mathrm{NN}}})^{0.3}$ behavior seen at lower energies 
\cite{Andronic:2004tx}, see Fig.~\ref{fig:dnchdy} (left panel). Shown 
are rapidity densities of charged particles $\ud N_{\mathrm{ch}}/\ud y$ at 
mid-rapidity, $y$=0 (where particles are emitted in the transverse direction); 
$y=\frac{1}{2}\ln\frac{E+p_{\mathrm L}}{E-p_{\mathrm L}}=
\mathrm{tanh}^{-1}(\beta_{\mathrm L})$, 
with $p_{\mathrm L}$ ($\beta_{\mathrm L}$) the longitudinal (beam direction) 
momentum (velocity), $E=\sqrt{m^2+p^2}$ the total energy.
The measurement at the LHC clearly demonstrated that the increase of 
$\ud N_{\mathrm{ch}}/\ud\eta$ with energy is steeper in nucleus-nucleus (AA) 
collisions compared to pp collisions, where the functional form is 
$(\sqrt{s_{\mathrm{NN}}})^{0.22}$ \cite{Aamodt:2010pb}.
The centrality dependence of $\ud N_{\mathrm{ch}}/\ud\eta$ is at the LHC 
\cite{Aamodt:2010cz} identical to that measured at RHIC, pointing out to 
a similar mechanism of particle production at the two energies.
A  model of the parton structure of matter at low parton fractional momentum
$x$, the Color Glass Condensate \cite{Gelis:2010nm}, is in a good agreement 
with the data
(see an extended comparison to theoretical models in \cite{Aamodt:2010cz}).
The data points shown in Fig.~\ref{fig:dnchdy} (left panel) are obtained 
by summing the measured $\ud N/\ud y$ yields for pions, kaons and protons 
and their antiparticles, see below.

Utilizing, in addition to particle counting, the momentum measurement 
(or, alternatively measuring the total hadron energy in calorimeters),
see Fig.~\ref{fig:dnchdy} (right),
one can extract from the data the energy density at the thermalization 
time. This involves a space-time model of the collision, which was put 
forward by Bjorken \cite{Bjorken:1982qr}. 
In this  model, the energy density is 
$\varepsilon = \frac{1}{A_{\rm T}}\frac{\ud E_{\rm T}}{\ud\eta}
\frac{\ud\eta}{\ud z}
= \frac{1}{A_{\rm T}}\frac{\ud E_{\rm T}}{\ud\eta}\frac{1}{c\tau_0}$,
where $E_{\mathrm T}$ is the transverse energy and $A_{T}=\pi R^2$ is the
geometric transverse area of the fireball (for central Pb--Pb collisions, 
$A_{T}\simeq$150~fm$^2$).
Assuming a conservative value for the equilibration time, $\tau_0$=1 fm/$c$,
one calculates for the LHC energy a matter energy density of
$\varepsilon_{LHC}$=15 GeV/fm$^3$ \cite{Chatrchyan:2012mb}, well
above the ``threshold'' value for quark-gluon matter, calculated in lattice 
QCD \cite{Karsch:2001cy}, of about 1 GeV/fm$^3$.

\begin{figure}[htb]
\begin{tabular}{lr} \begin{minipage}{.49\textwidth}
\centering\includegraphics[width=1.05\textwidth]{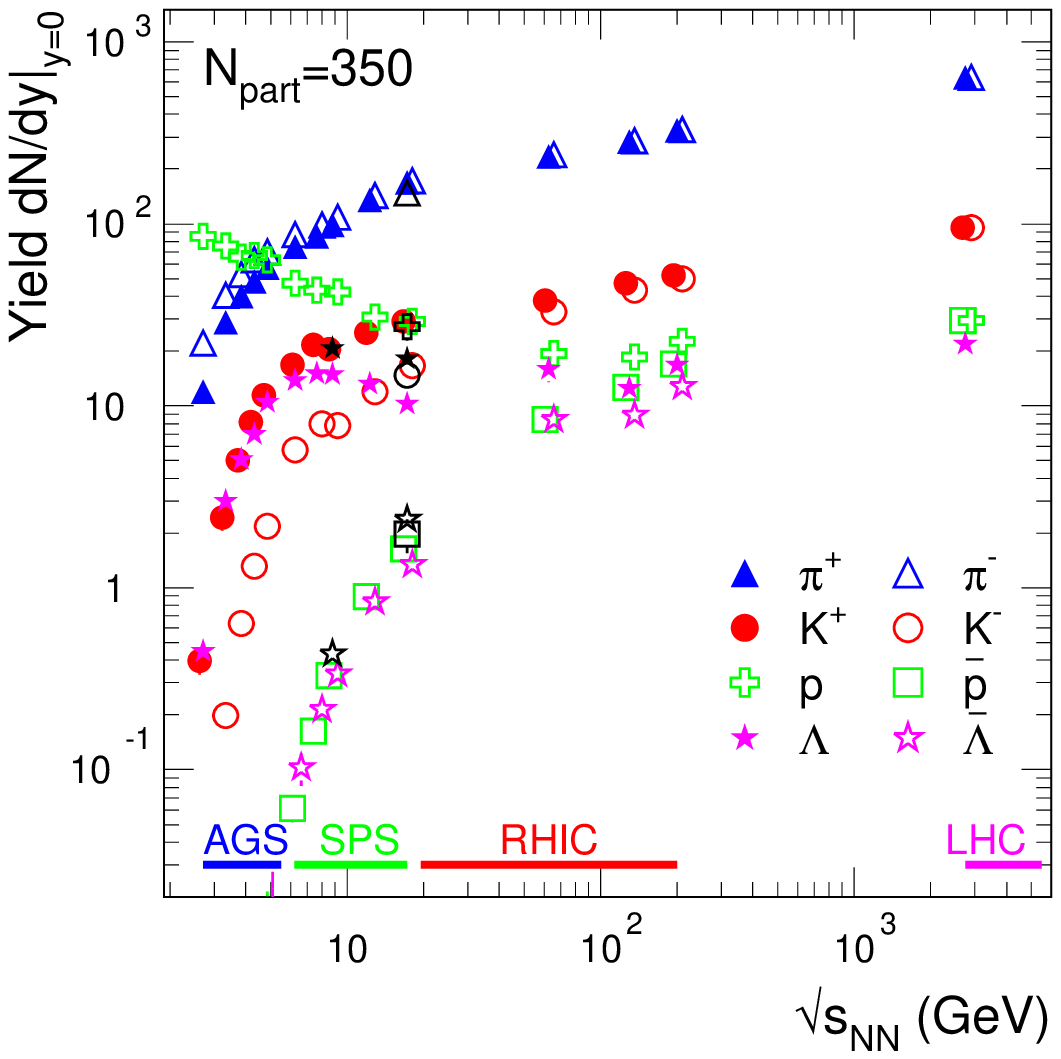}
\end{minipage} & \begin{minipage}{.49\textwidth}
\centering\includegraphics[width=1.05\textwidth]{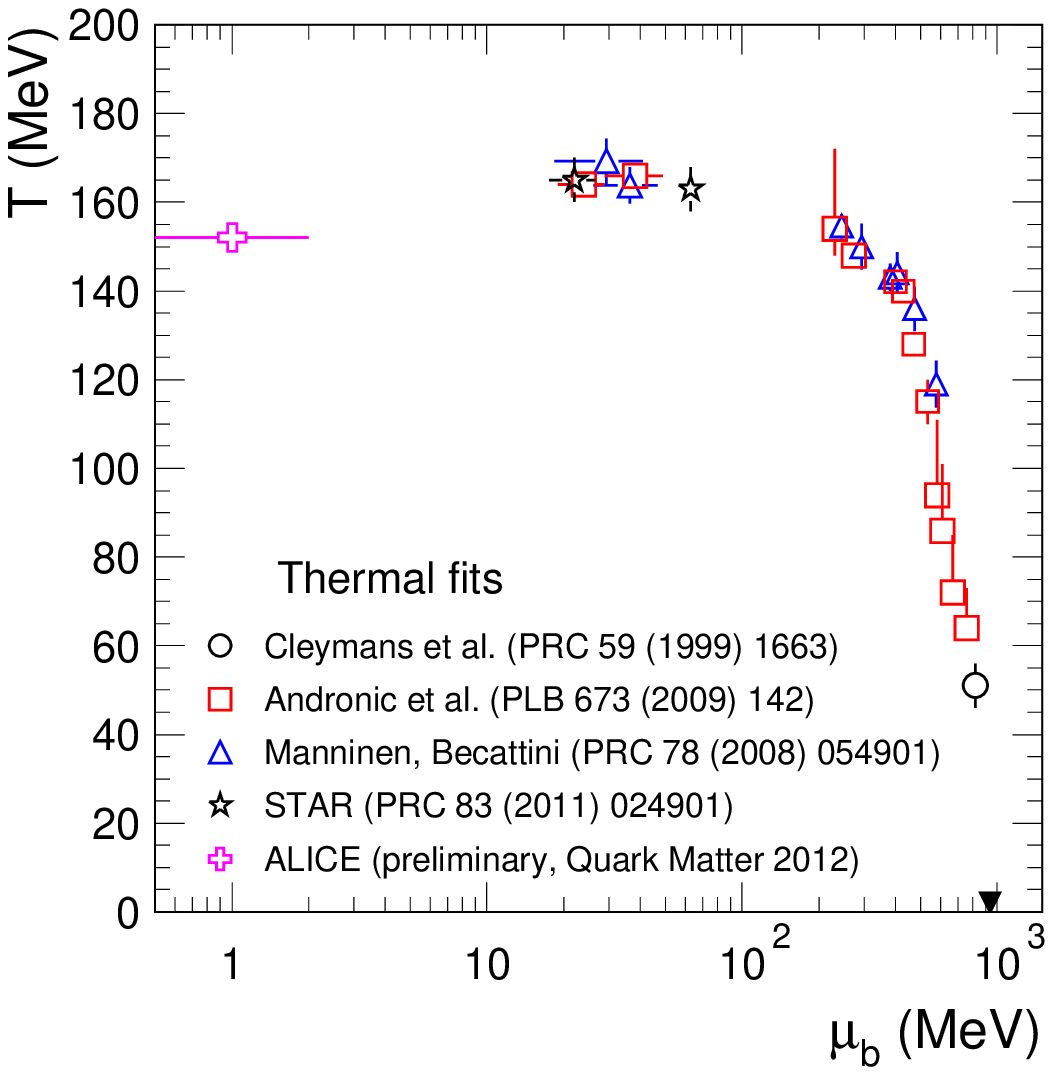}
\end{minipage}\end{tabular}
\caption{
Left panel: hadron multiplicities at mid-rapidity in central collisions.
Right panel: the phase diagram of strongly interacting matter with the points
representing the thermal fits of hadron yields. 
The down-pointing triangle indicates ground state nuclear matter (atomic 
nuclei).
} 
\label{fig:t-mu}
\end{figure}

In Fig.~\ref{fig:t-mu} (left panel) the collision energy dependence 
of identified hadron yields at mid-rapidity is shown. 
This comprises measurements by experiments at 
the AGS: E895 \cite{Klay:2001tf,Klay:2003zf,Pinkenburg:2001fj}, 
E866/E917 \cite{Ahle:1999uy,Ahle:2000wq}, E891 \cite{Ahmad:1998sg};
the SPS: NA49 \cite{Afanasiev:2002mx,Alt:2007aa,Alt:2005gr,Alt:2008qm}, 
NA44 \cite{Bearden:2002ib},
NA57 \cite{Antinori:2004ee};
RHIC: STAR \cite{Abelev:2008ab,Adler:2002uv,Adams:2006ke,Aggarwal:2010ig}, 
BRAHMS \cite{Arsene:2005mr}, 
PHENIX \cite{Adler:2003cb};
LHC: ALICE \cite{alice:2012iu}.
The monotonic decrease of the proton yield as a function of energy indicates
that fewer and fewer of the nucleons (or their valence $u,d$ quarks) in the 
colliding nuclei are ``stopped'' in the fireball. An onset of meson production 
is seen, with the kaons (heavier and containing a strange quark) produced 
less abundantly than pions.
The asymmetry between the $\pi^+$ and $\pi^-$ production yields reflects the 
isospin composition of the fireball. 
The asymmetry between $K^+$ and $K^-$ meson and $\Lambda$ and 
$\bar{\Lambda}$ hyperon production is determined by the quark content
of the hadrons: $K^+$($u\bar{s}$), $K^-$($\bar{u}s$)
$\Lambda$($uds$), $\bar{\Lambda}$($\bar{u}\bar{d}\bar{s}$).  
At lower energies, the availability in the fireball of valence $u$, $d$ 
quarks from colliding nucleons leads to a preferential production 
of hadrons carrying those quarks.
These asymmetries vanish gradually for higher 
energies, where the hadrons are mostly newly created (reflecting Einstein's 
famous equation $m=E/c^2$) and the production yields exhibit a clear mass 
ordering.
Good fits of the measurements are achieved with the thermal model with 3 
parameters: temperature $T$, baryochemical potential $\mu_b$, and volume $V$.
The thermal model describes a snapshot of the collision, namely the 
chemical freeze-out, which is assumed to be quasi-instantaneous.
This provides a phenomenological link of data to the QCD phase diagram, 
shown in Fig.~\ref{fig:t-mu}. Each point corresponds to a fit of hadron 
yields in central collisions of Au--Au or Pb--Pb nuclei at a given collision 
energy.
A remarkable outcome of these fits is that $T$ increases with increasing 
energy (decreasing $\mu_b$) from about 50 MeV to about 160 MeV, where
it exhibits a saturation for $\mu_b\lesssim$300 MeV.
This saturation of $T$ led to the connection to the QCD phase boundary,
via the conjecture that the chemical freeze-out temperature can be the 
hadronization temperature \cite{Andronic:2008gu,Andronic:2009gj}.
This picture is currently tested at the LHC with first identified hadron 
yields data \cite{alice:2012iu}.

\begin{figure}[hbt]
\begin{tabular}{lr} \begin{minipage}{.44\textwidth}
\centering\includegraphics[width=1.05\textwidth]{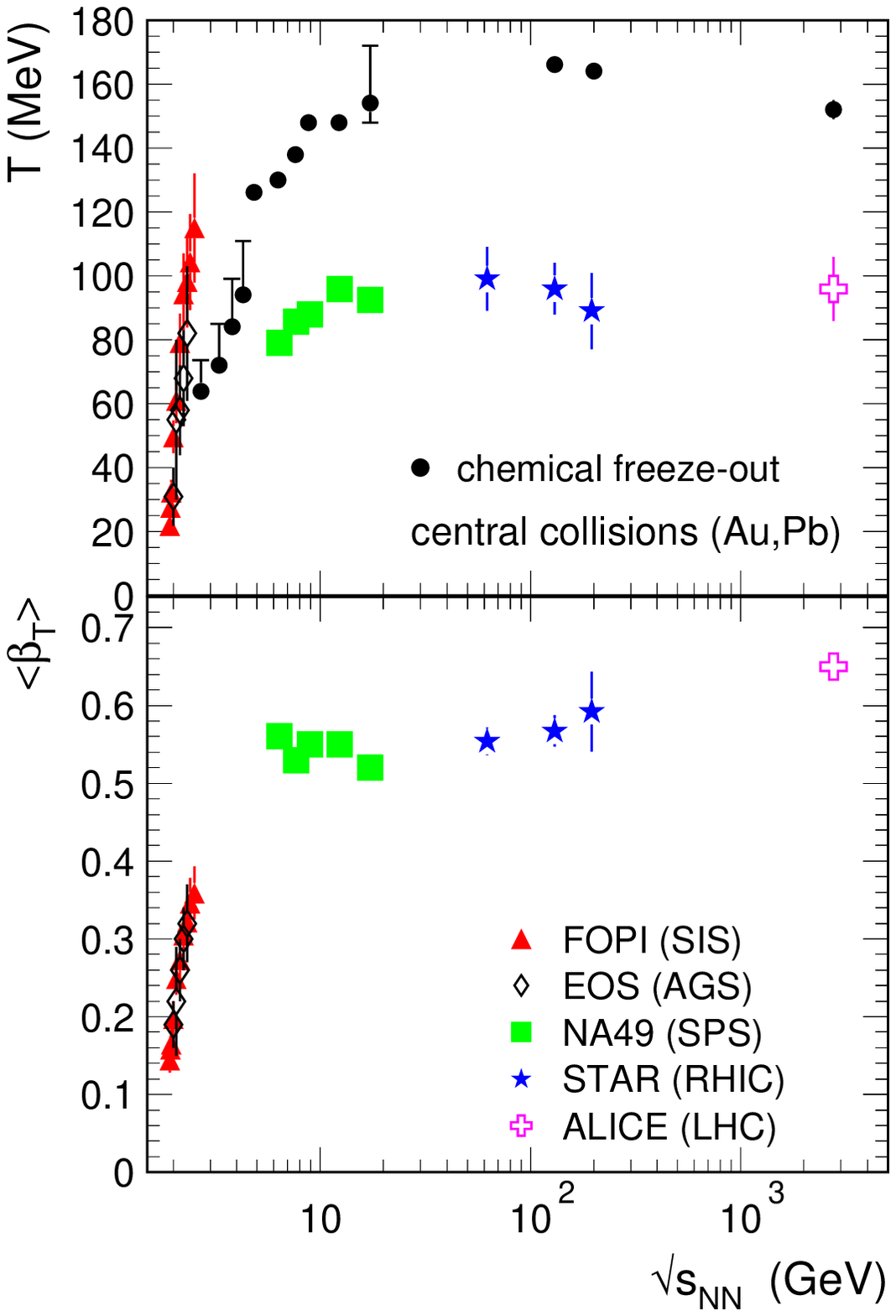}
\end{minipage} & \begin{minipage}{.59\textwidth}
\centering\includegraphics[width=.85\textwidth]{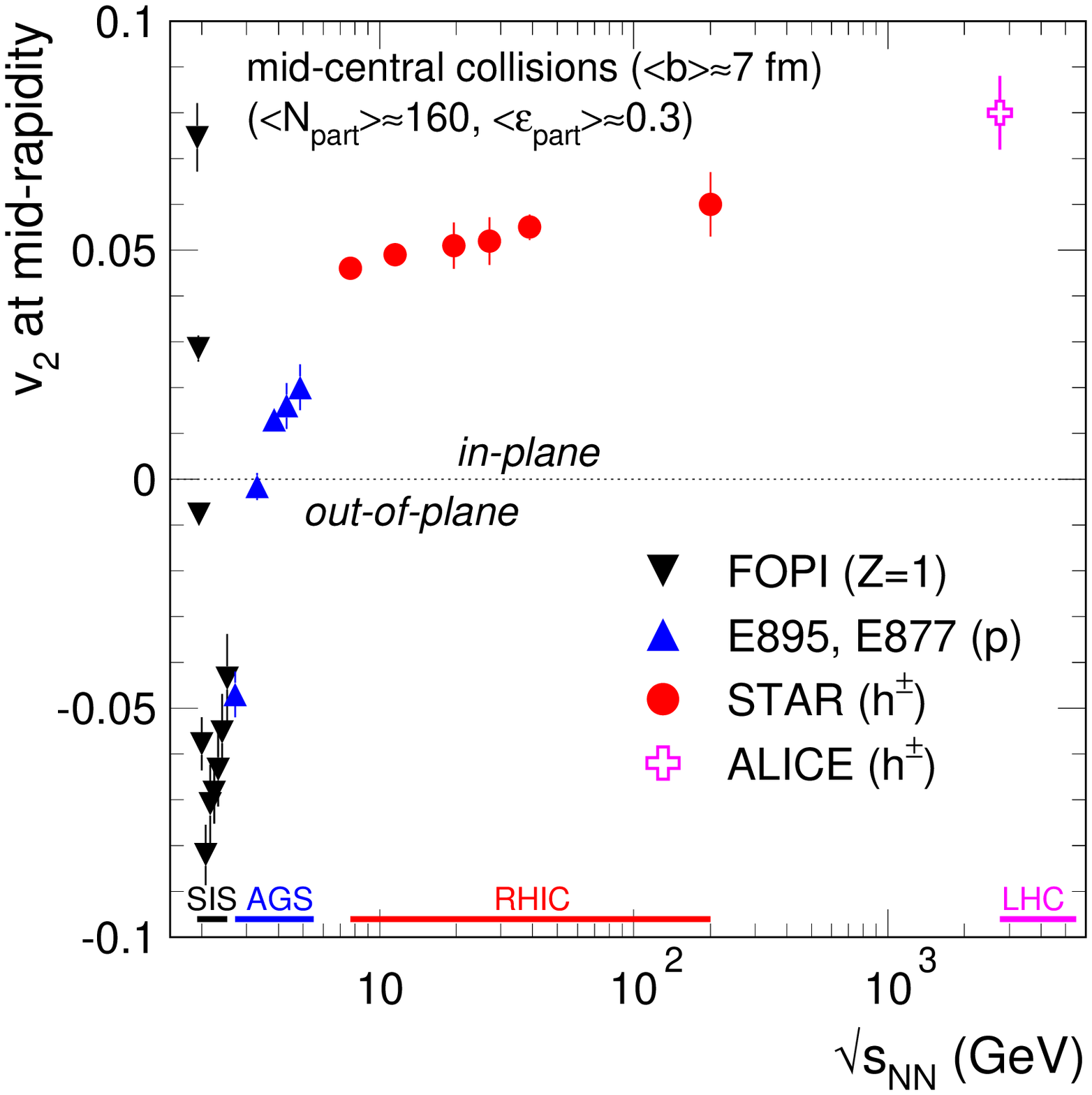}
\end{minipage}\end{tabular}
\caption{Collision energy dependence of collective flow: 
Left panel: radial flow in central collisions, quantified by the temperature
$T$ and average velocity $\langle \beta_{\mathrm T}\rangle$ at kinetic freeze-out; 
the temperature at chemical freeze-out is represented by the triangles.
Right panel: elliptic flow in mid-central collisions.}
\label{fig:flows}
\end{figure}

Collective flow is a distinct feature of nucleus-nucleus collisions.
In central collisions one investigates the so-called radial flow, which is 
quantified fitting transverse momentum ($p_{\mathrm T}$) spectra with 
the so-called ``blast wave'' model \cite{Schnedermann:1993ws}, accessing 
in a convenient (albeit maybe over-simplified) way bulk properties of the 
fireball at kinetic freeze-out.
The extracted fit parameters, the temperature $T$ and average transverse 
velocity $\langle \beta_{\mathrm T}\rangle$, are shown in Fig.~\ref{fig:flows} 
(left) as a function of the collision energy.
The measurements are by experiments 
FOPI \cite{Reisdorf:2010aa},
EOS \cite{Lisa:1994yr},
NA49 \cite{Alt:2007uj},
STAR \cite{Abelev:2008ab},
ALICE \cite{alice:2012iu}.
A strong increase of both $T$ and $\langle\beta_{\mathrm T}\rangle$ is seen 
at low energies (beam energies of up to 1 GeV/$A$ on fixed target) with 
a small further increase of $\langle\beta_{\mathrm T}\rangle$ and 
a constant kinetic freeze-out $T$, which is 50-60 MeV lower than the
chemical freeze-out $T$. At lower energies, the chemical freeze-out $T$ 
is smaller than the kinetic $T$, which is unphysical and awaits a resolution.
At the LHC, $\langle\beta_{\mathrm T}\rangle\simeq$0.65$c$ \cite{alice:2012iu}.

In Fig.~\ref{fig:flows} (right) we show the energy dependence of elliptic flow,
measured in mid-central collisions ($N_{part}\simeq 160$, corresponding to an
average impact parameter value of $\langle b\rangle\simeq 7$ fm) by experiments
FOPI \cite{Andronic:2004cp},
E895 \cite{Pinkenburg:1999ya}
E877 \cite{Barrette:1996rs},
STAR \cite{Adamczyk:2012ku},
ALICE \cite{Aamodt:2010pa}.
Elliptic flow arises in non-central collisions of nuclei as
a result of the initial elliptic trasverse shape of the overlap zone of 
the two nuclei (participant eccentricity $\varepsilon_{part}$). 
Through the initial gradients of the energy density (or pressure), this
leads to anisotropic spatial (angular) emission of hadrons.
This is quantified by the second order (quadrupole) Fourier coefficient
$v_2=\langle \cos(2\phi)\rangle$, where $\phi$ is the azimuthal angle with 
respect to the reaction plane.
The complex evolution of elliptic flow as a function of energy seen
in Fig.~\ref{fig:flows} (right) is understood qualitatively rather well.
At low energies, in-plane ($v_2>0$), rotation-like, emission may arise due
to low energy density in the overlap region and of long reaction times.
The fast transition towards preferential emission out-of-plane ($v_2<0$) 
is the outcome of more energetic collisions, leading to a larger
energy density of the fireball.
The increase of elliptic flow is a fingerprint of a stronger collective 
expansion, hindered by the passing spectators, which act as a shadow for 
the outgoing nucleons and fragments. The competition between the increasing 
speed of the expansion and of the decreasing passage time $t_{pass}$ of 
spectators leads to a maximum of (absolute value) elliptic flow in the SIS 
energy range. In this energy domain, the transiting spectators, with 
$t_{pass}$ varying between 40 and 10 fm/c, act as a clock for the collective 
expansion. In this regime, elliptic flow (historically called ``squeeze-out'')
is a prominent observable for the extraction of the nuclear EoS 
\cite{Danielewicz:2002pu}.
Towards larger energies, elliptic flow exhibits another transition, to
a preferential in-plane emission \cite{Ollitrault:1992bk}, a result of 
a unhindered collective expansion of the initially-anisotropic fireball.
Elliptic flow is built mostly in the earlier stages of the collision, 
since it is determined by the initial pressure gradients, 
which it alters quickly as it develops.
Consequently, at high energies, elliptic flow probes (albeit not 
exclusively) the deconfined state of quarks and gluons.

The long-awaited elliptic flow measurement at the LHC \cite{Aamodt:2010pa} 
exhibits a larger magnitude by $\simeq$35\% compared to the measurement at 
$\sqrt{s_{\mathrm{NN}}}=200$ GeV. This increase is described by hydrodynamics 
and was anticipated on a purely phenomenological $\log(\sqrt{s_{\mathrm{NN}}})$ 
behavior seen at lower energies \cite{Andronic:2004tx}.
The data show \cite{Aamodt:2010pa} that the increase is due exclusively to 
the increase of the average transverse momentum of the hadrons, while the 
$p_{\mathrm{T}}$ dependence of $v_2$ is identical at LHC to that measured 
at RHIC.

Another set of experimental observables from the kinetic freeze-out stage are
obtained from Hanbury Brown and Twiss (HBT) interferometry, which allows 
to measure to the size of (a region of) the fireball and its lifetime 
\cite{Aamodt:2011mr}.
Elliptic flow, radial flow and HBT interferometry are the observables used 
to extract, via comparisons to hydrodynamic calculations, the viscosity of 
deconfined matter (more precisely, shear viscosity divided by entropy density, 
$\eta/s$ \cite{Physics.2.88}).
The remarkable description of flow and HBT interferometry in hydrodynamic 
models observed at RHIC is further confirmed and extended with the 
measurements at the LHC.
The quantitative determination of $\eta/s$ is dependent on the initial 
energy density distributions, which are calculated according to 
either Glauber (binary collisions scaling) or Color Glass Condensate 
model.
It appears that deconfined matter is characterized by low values
of $\eta/s$; for an estimate for nuclear matter in mildly-excited nuclei 
see \cite{Auerbach:2009ba}.
Thermalization at a very early stage of the collision, at or below 1 fm/$c$, 
as needed in hydrodynamic calculations, is presently a challenge to 
theory \cite{Berges:2012ks}.

Can we identify an onset of deconfinement based on (bulk) hadronic observables
discussed above?  
The change, in the range $\sqrt{s_{\mathrm{NN}}}\simeq$5-10 GeV, of fireball
properties (see Fig.~\ref{fig:flows}) is a possible fingerprint,
but further experimental and theoretical support is needed to conclude.

Probing the deconfined matter in a more direct way is done with special
observables of the early stage. One category comprises thermal photons
and low-mass dileptons \cite{Gale:2012xq}, produced in the 
hot (deconfined) matter over its entire lifetime.
Measurements of thermal photons at RHIC have shown \cite{Adare:2008ab} 
that the temperature averaged over the lifetime of the fireball is larger 
than the chemical freeze-out $T$.
Another category of QGP probes is the so-called hard probes, namely 
processes characterized by an energy scale (quantified by the transverse mass
$m_{\mathrm T}=\sqrt{m_0^2+p_{\mathrm T}^2}$, where $m_0$ is the rest mass of the 
hadron) above several GeV (well above the temperature of the medium). 
Examples of such observables are hadrons at high $p_{\mathrm T}$ (or jets) 
and hadrons containing heavy (charm or bottom) quarks.
They are produced at early times in the collision ($t=1/m_{\mathrm T}$, with 
\mbox{1 fm} = 1/(0.2 GeV) in the system of units where $\hbar=c=1$ commonly 
used in high-energy physics).

Proposed by Bjorken in 1982 \cite{Bjorken:1982tu}, ``jet quenching'',
the extinction of jets (due to the energy loss of the parent parton) in QGP 
was for the first time observed at RHIC 
\cite{Arsene:2004fa,Adcox:2004mh,Back:2004je,Adams:2005dq} and is a subject 
of intense study at the LHC 
\cite{Aamodt:2010jd,CMS:2012aa,Abelev:2012eq,Aad:2012is}.
The usual method to quantify jet quenching is via the nuclear modification 
factor, defined as: 
$R_{\mathrm{AA}}=(\ud N_{{\mathrm AA}}/\ud y)/(N_{coll}\cdot\ud N_{\mathrm{pp}}/\ud y)$, 
where $\ud N/\ud y$ denotes the yield of a given observable measured in 
nucleus--nucleus (AA) or proton--proton (pp) collisions and $N_{coll}$ is 
the average number of nucleon-nucleon collisions over the given centrality 
interval of AA collisions;
$N_{coll}$ is calculated using the Glauber model \cite{Miller:2007ri}.

A change of physics in AA collisions (which in specialized terms is called 
a ``medium effect'') is seen as a departure of $R_{\mathrm{AA}}$ from unity.
However, modifications of parton distributions in nuclei compared 
to pp (shadowing or saturation) need to be considered carefully, in 
particular at LHC energies; very recent measurements in p--Pb collisions
address this issue \cite{ALICE:2012mj}.
We note that the binary collision scaling assumed in the construction of 
$R_{\mathrm{AA}}$ only applies to hard processes.
It is known experimentally that bulk particle production (comprising 
essentially pions, protons and kaons at low-momentum, 
$p_{\mathrm T}\lesssim$3-4 GeV/$c$) in AA collisions scales (in first order) 
with $N_{part}$ \cite{Aamodt:2010cz}.

\begin{figure}[hbt]
\begin{tabular}{lr} \begin{minipage}{.53\textwidth}
\centerline{\includegraphics[width=1.\textwidth]{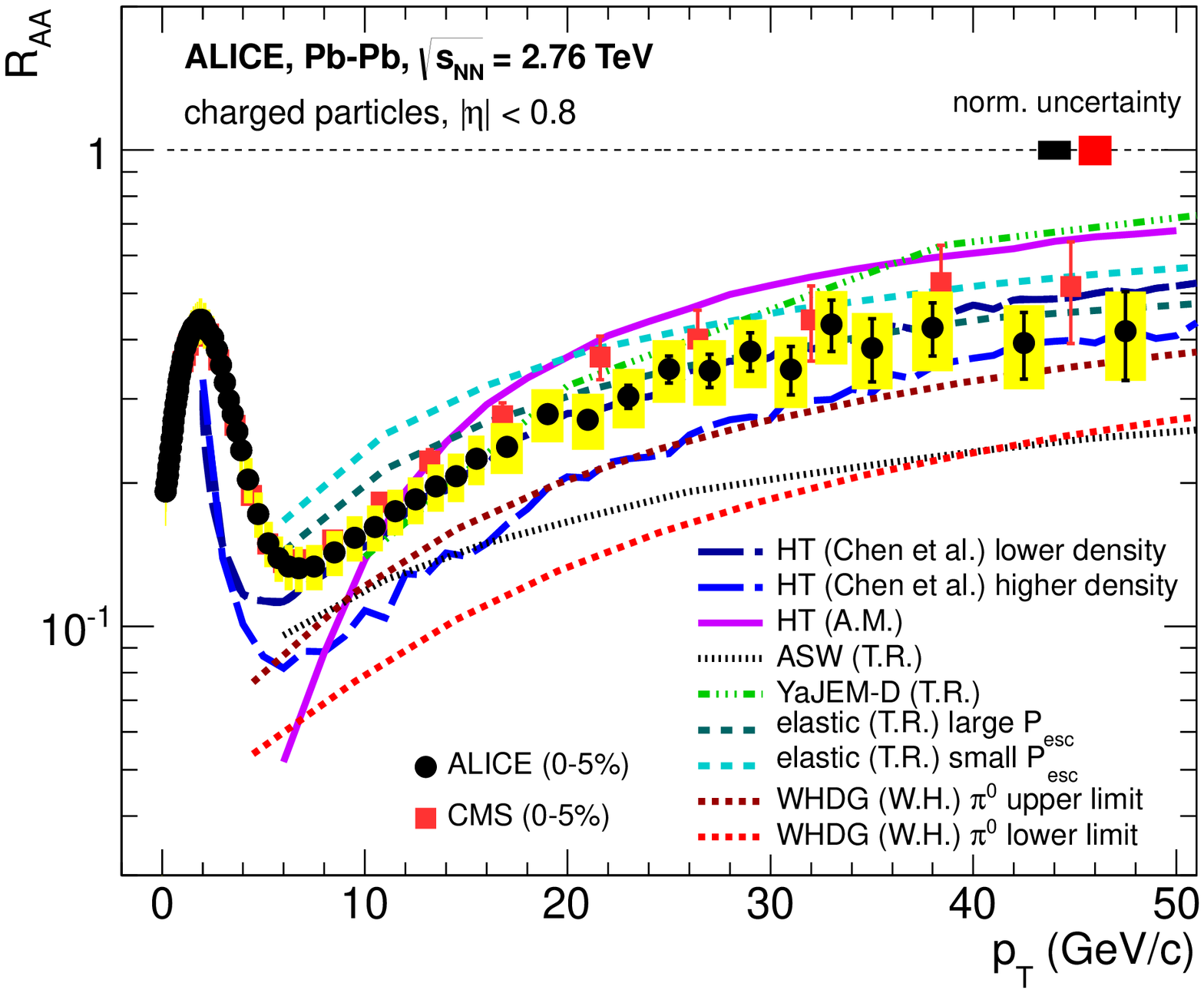}}
\end{minipage} & \begin{minipage}{.46\textwidth}
\centerline{\includegraphics[width=1.\textwidth]{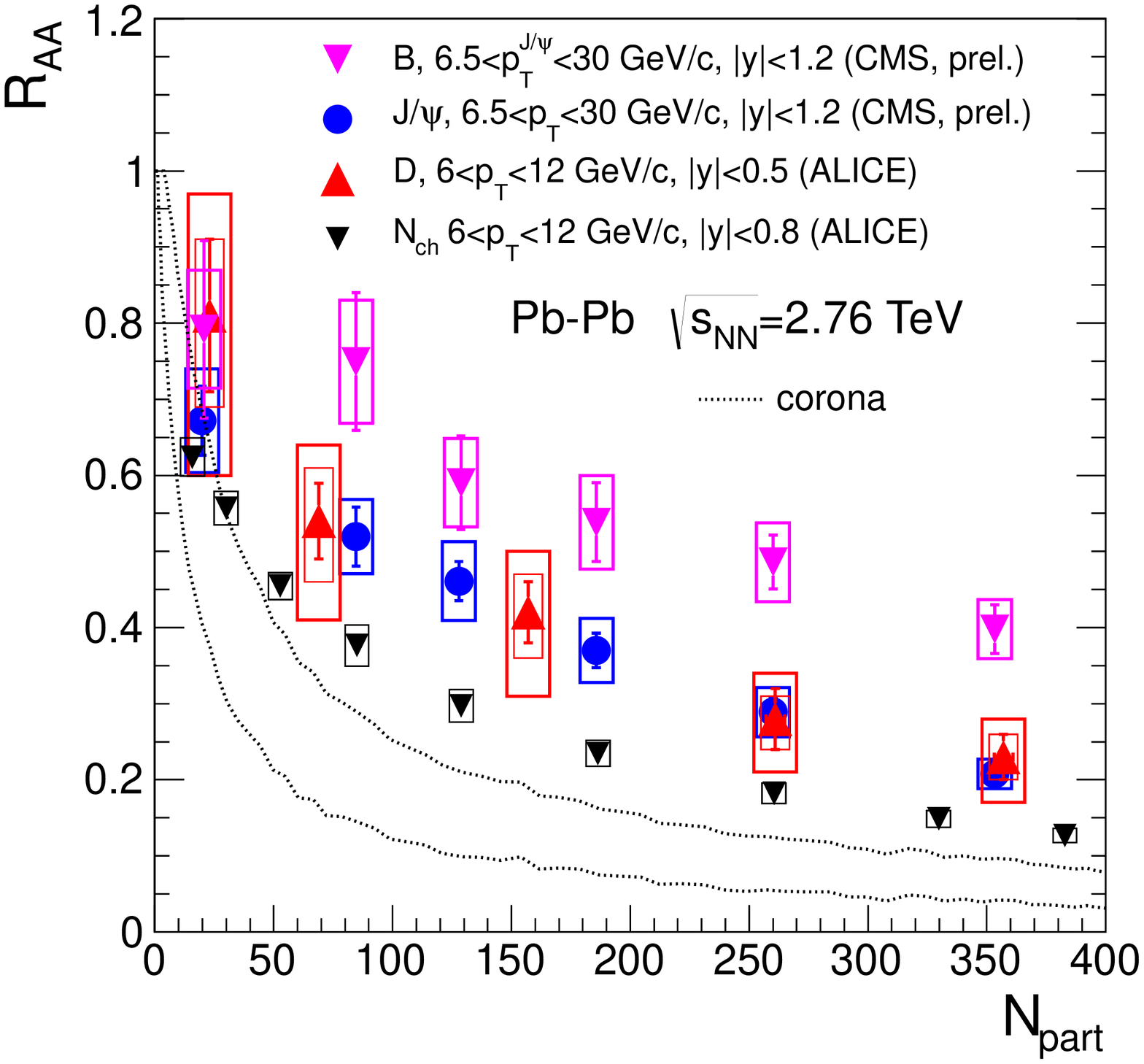}}
\end{minipage}\end{tabular}
\caption{Left panel: the nuclear modification factor $R_{\mathrm{AA}}$ as a 
function of transverse momentum for charged hadrons in Pb--Pb collsions at 
the LHC (Fig. from \cite{Abelev:2012eq}, see refs. therein for the theoretical
curves).
Right panel: centrality dependence of $R_{\mathrm{AA}}$ for charged hadrons at 
high $p_{\mathrm T}$ in comparison to that to charmed mesons ($D$),
charmonium (J/$\psi$) and beauty hadrons ($B$). 
The dotted lines denote the contribution of production in the corona of 
colliding nuclei.
} 
\label{fig:raanch} 
\end{figure}

The first measurement of $R_{\mathrm{AA}}$ for charged hadrons at the LHC 
\cite{Aamodt:2010jd} showed that the suppression is larger than 
previously measured at RHIC, reaching a factor of about 7. 
The suppression is reduced towards larger $p_{\mathrm T}$ values, 
see Fig.~\ref{fig:raanch} (left), but remains substantial even at 
50-100 GeV/$c$ \cite{CMS:2012aa,Abelev:2012eq};
it is also seen in reconstructed jets \cite{Aad:2012is}.

The basic features seen in the data are reproduced by models implementing
parton energy loss in deconfined matter. Towards extraction of transport 
coefficients, the description of jet quenching in theoretical models remains 
a challenging task \cite{Renk:2012wi}.
The measurements, in conjunction with theoretical models, clearly demonstrate
that partons lose energy in the hot and dense quark-gluon matter, which
leads to broader jets \cite{Chatrchyan:2012gw}.

For the first time measured at the LHC in a direct way, the nuclear 
suppression of charmed \cite{ALICE:2012ab} and beauty \cite{CMS-PAS-HIN-12-014}
hadrons at high $p_{\mathrm T}$ is shown as a function of centrality in 
Fig.~\ref{fig:raanch} (right). 
The theoretical expectation is that heavy quarks (charm and bottom) lose 
less energy (by gluon radiation) compared  to lighter (up, down, strange) 
ones \cite{Dokshitzer:2001zm}. 
This expectation is exhibited by the data, see Fig.~\ref{fig:raanch} (right),
although a definite conclusion needs further support from experiment.
The energy loss suffered by energetic heavy quarks in QGP 
is indicative of their ``strong coupling'' with the medium, dominated by 
light quarks and gluons.
The measurements at the LHC consolidate and extend the observation at RHIC 
of heavy quark energy loss and flow \cite{Adare:2010de}.

\begin{figure}[hbt]
\begin{tabular}{lr} \begin{minipage}{.42\textwidth}
\centering\includegraphics[width=1.02\textwidth]{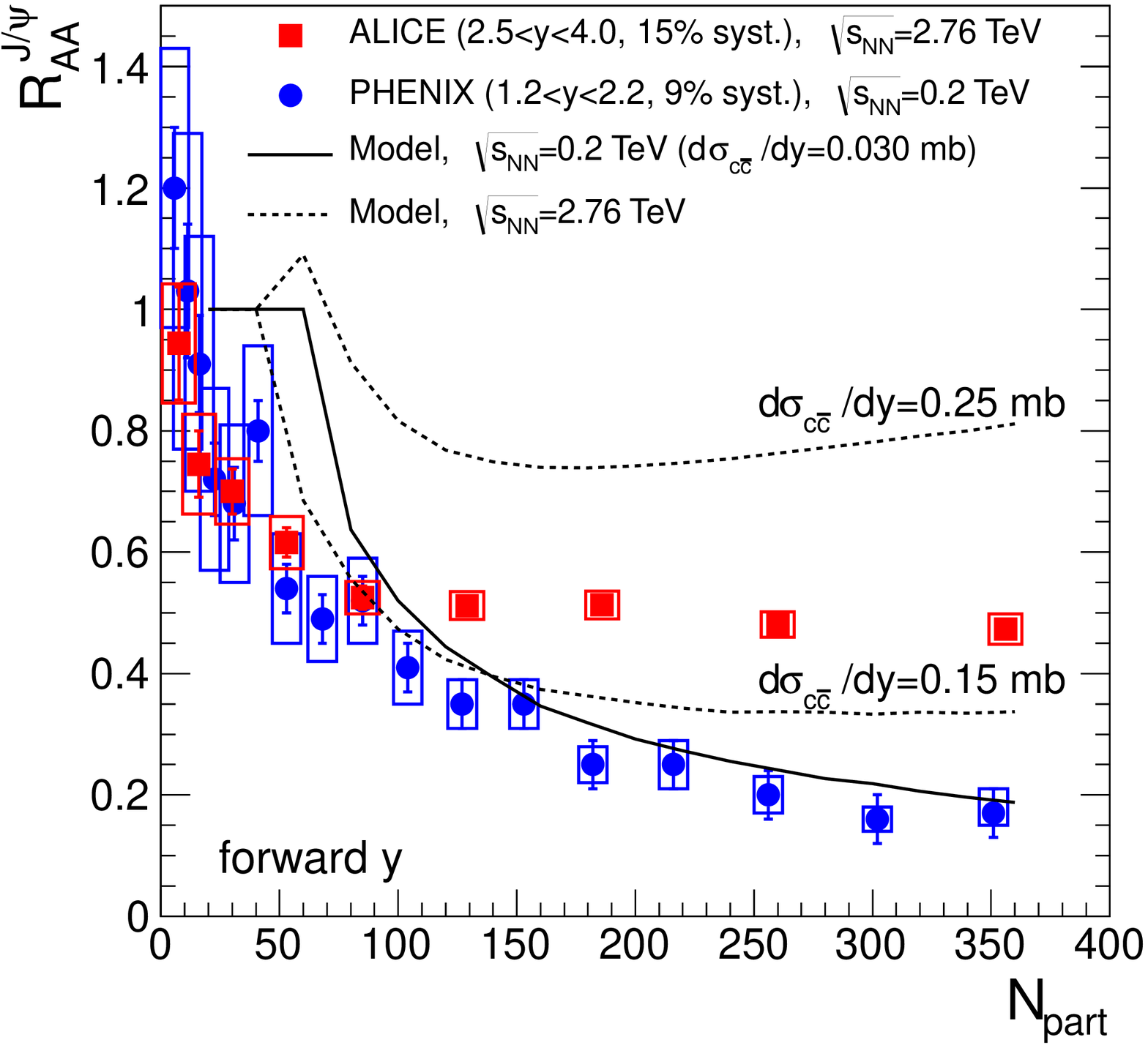}
\end{minipage} & \begin{minipage}{.56\textwidth}
\centering\includegraphics[width=1.\textwidth]{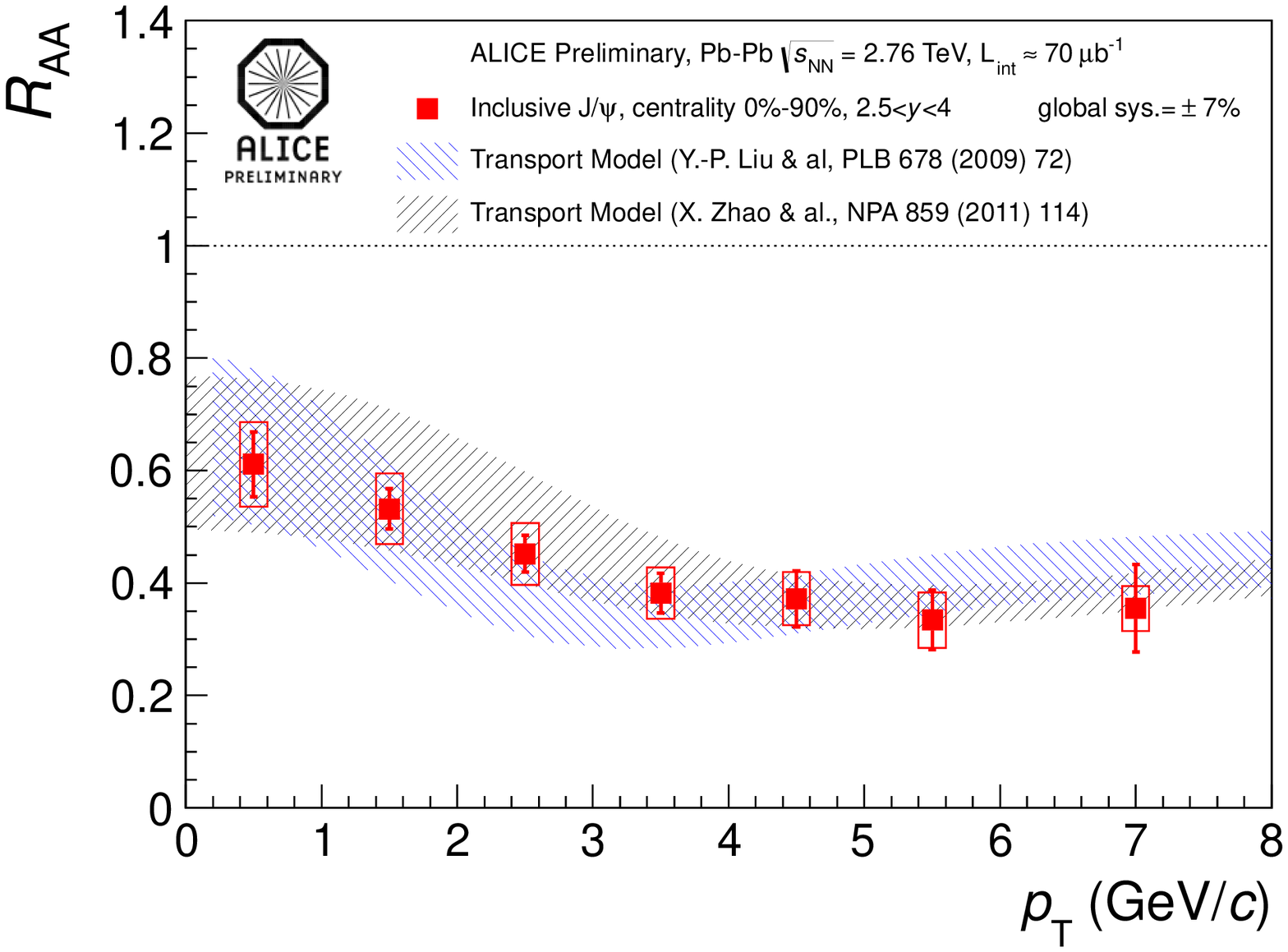}
\end{minipage}\end{tabular}
\caption{Left panel: centrality dependence of the nuclear modification factor 
for inclusive J$/\psi$ productions at RHIC and LHC energies. Data are 
compared to statistical hadronization model calculations.
Right panel: transverse momentum dependence of the nuclear modification 
factor of J/$\psi$ measured at the LHC by ALICE (preliminary data) in 
comparison with transport model calculations (plot from \cite{Suire:2012gt}).
} 
\label{fig:raajpsi}
\end{figure}

Among the various suggested probes of deconfinement, charmonium ($c\bar{c}$) 
states plays a distinctive role. J/$\psi$ is the first hadron for which 
a clear mechanism of suppression (melting) in QGP was proposed early on, 
based on the color analogue of Debye screening \cite{Matsui:1986dk}.
In the statistical hadronization model \cite{BraunMunzinger:2000px}, the 
charm quarks produced in initial hard collisions thermalize in QGP and 
are ``distributed'' into hadrons at chemical freeze-out. 
All charmonium states are assumed to be not formed at all in the deconfined 
state but are produced, together with all other hadrons, at chemical freeze-out.
See \cite{Andronic:2011yq} for recent predictions of this model.
Kinetic recombination of charm and anti-charm quarks in QGP \cite{Thews:2000rj}
is an alternative quarkonium production mechanism.
In this model (see \cite{Liu:2009nb,Zhao:2011cv} for recent results), 
a continuous dissociation and regeneration of charmonium takes place in the 
QGP over its entire lifetime. 

The measurement of J/$\psi$ production in Pb--Pb collisions at the LHC was 
expected to provide a definitive answer on the question of (re)generation.
The data measured at high-$p_{\mathrm T}$ \cite{Chatrchyan:2012np} show a 
pronounced suppression of J/$\psi$ in Pb--Pb compared to pp collisions
and of the same magnitude as that of open-charm hadrons, see 
Fig.~\ref{fig:raanch} (right).
This may indicate that the high-$p_{\mathrm T}$ charm quarks that form either 
$D$ or J/$\psi$ mesons had the same dynamics, possibly a thermalization 
in QGP and a late hadronization.

The first LHC measurement of the overall (inclusive in $p_{\mathrm T}$) 
production \cite{Abelev:2012rv}, showed $R_{\mathrm{AA}}$ values significantly 
larger than at RHIC energies, see Fig.~\ref{fig:raajpsi} (left). 
The data are well described by both statistical hadronization model 
\cite{Andronic:2011yq}, shown in Fig.~\ref{fig:raajpsi} (left), 
and by transport models \cite{Liu:2009nb,Zhao:2011cv}, which also describe
the $p_{\mathrm T}$ dependence of $R_{\mathrm{AA}}$ \cite{Suire:2012gt}, 
Fig.~\ref{fig:raajpsi} (right). The transport model calculations show that, 
as expected, that regeneration is predominatly a low-$p_{\mathrm T}$ phenomenon.
Discriminating the two pictures, statistical production at chemical freeze-out
and regeneration during the QGP lifetime, may help providing an answer to 
fundamental questions related to the fate of hadrons in a hot medium \cite{Laine:2011xr}.
Recent measurements at the LHC of production of bottonium ($b\bar{b}$)
states \cite{CMS:2012fr} adds one more flavor to the QGP research field.

\bibliographystyle{aipproc}   

\bibliography{qgp}

\IfFileExists{\jobname.bbl}{}
 {\typeout{}
  \typeout{******************************************}
  \typeout{** Please run "bibtex \jobname" to optain}
  \typeout{** the bibliography and then re-run LaTeX}
  \typeout{** twice to fix the references!}
  \typeout{******************************************}
  \typeout{}
 }

\end{document}